\documentstyle[12pt]{article}
\newcommand{\slp}{\raise.15ex\hbox{$/$}\kern-.57em\hbox{$\partial$}}
\newcommand{\sla}{\raise.15ex\hbox{$/$}\kern-.57em\hbox{$a$}}
\newcommand{\slA}{\raise.15ex\hbox{$/$}\kern-.57em\hbox{$A$}}
\newcommand{\slb}{\raise.15ex\hbox{$/$}\kern-.57em\hbox{$b$}}
\begin{document}
\def\sqr#1#2{{\vcenter{\hrule height.#2pt
   \hbox{\vrule width.#2pt height#1pt \kern#1pt
    \vrule width.#2pt}
    \hrule height.#2pt}}}
\def\dal{\mathchoice{\sqr{6}{4}}{\sqr{5}{3}}{\sqr{5}3}{\sqr{4}3} \, }
\renewcommand{\Box}{\dal}
\def\until{\lower4pt\hbox{$\sim $}  \! \! \! \! }
\def\lflex{\raise 6pt\hbox{$\leftarrow $}\! \! \! \! \! \! }
\def\be{\begin{equation}}
\def\ee{\end{equation}}
\def\beq{\begin{equation}}
\def\eeq{\end{equation}}
\def\bea{\begin{eqnarray}}
\def\eea{\end{eqnarray}}
\def\bear{\begin{eqnarray}}
\def\eear{\end{eqnarray}}
\def\bearst{\begin{eqnarray*}}
\def\eearst{\end{eqnarray*}}
\def\ear{\end{eqnarray}}
\def\ba{\begin{eqnarray*}}
\def\ea{\end{eqnarray*}}
\def\peleven{\parbox{11cm}}
\def\peffec{\peight{\bearst\eearst}\hfill\peleven}
\def\peffsev{\pseven{\bearst\eearst}\hfill\pnine}
\def\pspace{\peight{\bearst\eearst}\hfill}
\def\pspacem{\pseven{\bearst\eearst}\hfill}
\def\ptwelve{\parbox{12cm}}
\def\peight{\parbox{8mm}}
\def\pseven{\parbox{7mm}}
\def\pnine{\parbox{9cm}}
\def\E{{e}}
\newcommand\pa{\partial}
\newcommand\un{\underline}
\newcommand\ti{\tilde}
\newcommand\pr{\prime}
\newcommand{\cl}{{\cal{cl}}}
\def\cP{\cal P}
\def\cH{\cal H}
\def\cL{\cal L}
\def\eff{\cal {eff}}
\def\Li{\cal Li}
\def\D{\cal D}
\def\cD{\cal D}
\def\cB{\cal B}
\def\half{\frac{1}{2}}
\def\undertilde#1{{\baselineskip=0pt\vtop{\hbox{$#1$}
\hbox{$\scriptscriptstyle\sim$}}}{}}
\newcommand{\eins}{  1\!{\rm l}  }
\newcommand{\dA}{sqcup\!\!\!\!\!\sqcap}
\renewcommand{\thefootnote}{\fnsymbol{footnote}}
\font\grrm=cmbx10 scaled 1200
\font\vb=cmbx10 scaled 1440
\font\bigcal=cmsy10 scaled 1200
\def\eightpoint{\def\rm{\fam0\eightrm}}
\def\flex{\raise 6pt\hbox{$\leftrightarrow $}\! \! \! \! \! \! }
\def\tr{ \mathop{\rm tr}}
\def\atanh{\mathop{\rm atanh}}
\def\Tr{\mathop{\rm Tr}}
 \def\smin{\,\raise 0.06em \hbox{${\scriptstyle \in}$}\,}
\def\smsubset{\,\raise 0.06em \hbox{${\scriptstyle \subset}$}\,}
\def\smwedge{\,{\scriptstyle \wedge}\,}
\def\smvee{\,{\scriptstyle \vee}\,}

\def\Natural{\hbox{\hskip 1.5pt\hbox to 0pt{\hskip -2pt I\hss}N}}
\def\Integer{\>\hbox{{\sf Z}} \hskip -0.82em \hbox{{\sf Z}}\,}
\def\Rational{\hbox{\hbox to 0pt{\hskip 2.7pt \vrule height 6.5pt
                                  depth -0.2pt width 0.8pt \hss}Q}}
                                  
\def\Real{\hbox{\hskip 1.5pt\hbox to 0pt{\hskip -2pt I\hss}R}}
\def\Complex{\hbox{\hbox to 0pt{\hskip 2.7pt \vrule height 6.5pt
                                  depth -0.2pt width 0.8pt \hss}C}}
\def \ln {{\rm ln}\, }
\def \cotg {\rm cotg } 
\input epsf.tex
\newcommand{\LL}{{\cal{L}}}
\begin{titlepage}
{\tt HD-THEP-07-13}
\begin{center}
{\large{\bf From the BRST invariant Hamiltonian to the Field-Antifield Formalism}}\par
\vskip 0.3cm
\end{center}
\begin{center}
{Heinz J. Rothe and Klaus D. Rothe}\par
\vskip 0.3cm
{Institut f\"ur Theoretische Physik}\par
{Universit\"at Heidelberg,\\
Philosophenweg 16, D-69120 Heidelberg, Germany}
\footnote{email: h.rothe@thphys.uni-heidelberg.de\\
k.rothe@thphys.uni-heidelberg.de}\\
{\today}
\end{center}

\begin{abstract}
\noindent
We study the relation between the lagrangian field-antifield formalism and the BRST invariant phase space formulation of gauge theories. 
Starting from the Batalin-Fradkin-Vilkovisky unitarized action, we demonstrate in a 
deductive way the equivalence of the phase space, and the lagrangian field-antifield  
partition functions for the case of irreducible first rank theories. 

\bigskip\noindent
PACS: 11:10Ef; 11:15-q
\end{abstract}
\end{titlepage}

\section{Introduction}

The phase-space representation of the quantum 
partition function for a gauge field theory suffers from two main drawbacks: i) one generally wants to formulate the Feynman rules in configuration space, and ii) the hamiltonian formulation is not manifestly covariant, so that it is not clear which gauge fixing condition in phase space implies a covariant choice of gauge in the lagrangian formulation.

The field-antifield  approach of Batalin and Vilkovisky (BV) to the quantization of theories with a local symmetry solves both of these problems [1-5].
In this formulation the solution of a so called master equation provides the configuration-space counterpart of the Batalin-Fradkin-Vilkovisky (BFV) phase-space quantum action [6-10].

The field-antifield quantization is usually presented in an axiomatic way, 
its a priori justification relying ultimately on the observation that the action, determined from a master equation,
can be shown to embody the full gauge structure of the theory in question [9,11]. In this approach one departs from the classical configuration-space action $S_{cl}[q]$ as a functional of the ``fields" $q^i, i=1,\cdots, n$, assumed to be invariant under an $r$ parameter group of gauge transformations.
For every gauge parameter $\varepsilon^\alpha(t)$ one introduces a ghost field $c^\alpha(t)$,
and with each of the fields $(q^i,c^\alpha)$ one associates a corresponding ``antifield" $(q^*_i,c^*_\alpha)$.

The relation of the field-antifield approach to the BFV hamiltonian approach has been discussed from various points of view by starting either from the axiomatic BV-lagrangian formulation, or from the BFV-hamiltonian formulation [12-24]. The equivalence of the two formalisms for arbitrary gauge systems has however not yet been proven, nor does this paper provide a general proof. 
The purpose of this paper is to elucidate in a simple and clear way the connection between the two approaches to quantization, and to show that for first rank theories with a closed algebra the hamiltonian BFV formulation, which is well understood, leads in a very natural and straight forward way to the lagrangian quantization of Batalin and Vilkovisky.
Our deductive procedure will shed light on i) the role played by the antifields, and 
the connection between the coefficient functions, multiplying these fields in the action, to pull-backs of the 
BRST variations that implement the symmetry of the quantum action in phase space; 
ii) the non-minimal extension of the minimal solution to the 
master equation in the lagrangian approach [22,24], which is automatically 
generated by the BRST 
exact contribution in the hamiltonian formulation, and iii) the connection between the lagrangian master equation, and the hamiltonian master equation. 

The paper is organized as follows: in the following section we consider the well known BRST hamiltonian formulation [6-10] and introduce, via the BRST exact contribution, a set of hamiltonian antifields. The corresponding action solves a hamiltonian master equation without any restrictions on the gauge theory considered.
In section 3 we then consider the transition to configuration space. At this point we shall concentrate on theories of rank 1 with a gauge algebra whose structure functions do not depend on the momenta. This allows us to perform 
the momentum integrations in order to arrive at a generic form for the field-antifield action. 
One is then led in a straight forward way to the field-antifield action of Batalin and Vilkovisky with a "non minimal extension".

\section{From the BFV-phase-space action \\
to the hamiltonian master equation}
Our starting point is the BRST invariant phase-space action  
of Batalin, Fradkin and Vilkovisky [6-10] for pure first class systems, which allows one to quantize also gauge theories that cannot be handled by the Faddeev-Popov trick.
This action is determined once the BRST invariant Hamiltonian $H_{B}$ and the nilpotent BRST charge $Q_{\cB}$ are given.
Both quantities can be constructed in a systematic way from the knowledge of the
involutive
Poisson algebra of the first class constraints $G_A$ with themselves, and with the canonical Hamiltonian $H_c$ [8]
\be
\{G_A,G_B\} = G_C U^C_{AB} \ ,\label{13.GG-PB}
\ee
\be
\{G_A,H_c\} = G_B V^B_{\,\,A}\ .\label{13.HG-PB}
\ee
Here $U^C_{AB}$ and $V^B_{\ A}$ can depend on the coordinates $q^i$ and their conjugate momenta $p_i$. 
The ``unitarized" BRST invariant phase space action is then given by 
\be\label{13.BV-action}
S_{U}[q,p,\eta,\bar{\cal P}]=\int dt \Bigl(\dot q^ip_i + \dot\eta^A{\bar{\cP}}_A 
- {H}_{B}(q,p,\eta,\bar{\cal P})+ \{\Psi,Q_{\cB}\}\Bigr)\,,
\ee
where $\eta^A$ and ${\bar{\cal P}}_A$ are canonically conjugate Grassmann valued pairs carrying ghost number 1 and -1, respectively.
$\Psi$ is an arbitrary ``fermion gauge fixing function" in phase-space, carrying Grassmann signature $\epsilon(\Psi) = 1$ and ghost number $gh(\Psi) = -1$.
The contribution $\{\Psi,Q_{\cB}\}$ is a BRST exact term, with the 
generalized (graded) Poisson brackets of two functions of the canonical variables $Q^k$ and $P_k$ defined as follows 
\footnote{With this definition the algebraic properties of the Poisson brackets are those of the Berezin algebra, and the equations of motion take the standard form. See e.g. [25].}
\be\label{13.PB1}
\{F,G\} = -\sum_k\left(\frac{\partial^{(r)}F}{\partial P_k} 
\frac{\partial^{(\ell)}G}{\partial Q^k} -(-1)^{\epsilon(Q)\epsilon(P)}
\frac{\partial^{(r)}F}{\partial Q^k} 
\frac{\partial^{(\ell)}G}{\partial P_k}\right)\,,
\ee
where $\partial^{(r)}$ ($\partial^{(\ell)}$) denotes the right (left) derivative, and $\epsilon(f)$ is the Grassmann signature of $f$. Note that (\ref{13.PB1}) implies that
\[
\{\eta^A,{\bar{\cal P}}_B\} = -\delta^A_{\ B}\ .
\]
We can then rewrite the $BRST$ exact contribution in (\ref{13.BV-action}) as follows:
\[
\{\Psi,Q_{\cB}\} = \sum_i\left(\frac{\partial\Psi}{\partial q^i}\frac{\partial Q_{\cB}}{\partial p_i} - 
\frac{\partial\Psi}{\partial p_i}\frac{\partial Q_{\cB}}{\partial q^i}\right)
-\sum_A\left(\frac{\partial^{(r)}\Psi}{\partial\eta^A}
\frac{\partial^{(\ell)}Q_{\cB}}{\partial {\bar{\cP}}_A} + 
\frac{\partial^{(r)}\Psi}{\partial {\bar{\cP}}_A}\frac{\partial^{(\ell)} Q_{\cB}}{\partial \eta^A}\right)\ .
\] 
Alternatively this expression can be written in the form
\bearst
\{\Psi,Q_{\cB}\} &=& -\sum_i\left(\frac{\partial\Psi}{\partial q^i}\delta_B q^i + 
\frac{\partial\Psi}{\partial p_i}\delta_B p_i\right)\\
&+&\sum_A\left(\frac{\partial^{(r)}\Psi}{\partial \eta^A}\delta_B \eta^A + 
\frac{\partial^{(r)}\Psi}{\partial {\bar{\cP}_A}}\delta_B {\bar{\cP}}^A\right)\,,
\eearst
where $\delta_B f$ is the BRST variation of $f$ defined by
\footnote{Note the ordering of the entries in the graded Poisson bracket. This is consistent with 
the definition of the BRST variation with the "fermionic" $\hat s$ operator: $\hat s(fg) = (\hat s f)g + (-1)^{\epsilon(f)}f(\hat s g).$}
\be\label{13.delta_B}
\delta_B f = \{Q_{\cB},f\}\ .
\ee
$\Psi$ is a completely arbitrary function of the canonical variables with Grassmann parity ${\epsilon(\Psi)=1}$ and ghost number $gh(\Psi)=-1$, which reflects the arbitraryness in the choice of gauge. Let us therefore introduce the following partial derivatives of $\Psi$ as new variables:
\bear\label{13.def.antifield(1)}
q^*_i &=& -\frac{\partial\Psi}{\partial q_i}\,,\quad
p^{*i} = -\frac{\partial\Psi}{\partial p_i}\nonumber\\
\eta^*_A &=& \frac{\partial^{(r)}\Psi}{\partial \eta^A}\,,\quad
\bar{\cP}^{*A} = \frac{\partial^{(r)}\Psi}{\partial{\bar{\cP}}_A}\,.
\eear
The above derivatives are functions of $q^i$, $p_i$, $\eta^A$ and ${\bar{\cP}}_A$. For later convenience it is useful to also introduce the functional
\[
\Psi_H[q, p, \eta,{\bar{\cP}}]  = \int dt \,\Psi(q(t), p(t), \eta(t),{\bar{\cP}}(t))\,.
\]
Then (\ref{13.def.antifield(1)}) can be rewitten in terms of functional derivatives as follows:
\bear\label{13.def.antifield(2)}
q^*(t) &=& -\frac{\delta\Psi_H}{\delta q^i(t)}\,,\quad
{p^*}^i(t) = -\frac{\delta\Psi_H}{\delta p_i(t)}\nonumber\\
\eta^*_A(t) &=& \frac{\delta^{(r)}\Psi_H}{\delta \eta^A}\,,\quad
\bar{\cP}^{*A}(t) = \frac{\delta^{(r)}\Psi_H}{\delta{\bar{\cP}}_A(t)}\,.
\eear
We call the ``star" variables "phase-space antifields".

Let us denote the variables collectively by $\{\theta^\ell\}
\equiv \{q^i, p_i,\eta^A, {\bar{\cal P}}_A\}$ and $\{\theta^*_{\ell}\} \equiv 
\{q^*_i, p^{* i}, \eta^*_A, {\bar{\cal P}}^{* A}\}$.
Since $\Psi_H$ carries Grassmann 
signature and ghost number 1 and -1, respectively, the
Grassmann signature and ghost number of a ``star"-variable $\theta^*$ is given according to 
(\ref{13.def.antifield(2)}) by
\bearst
&&\epsilon(\theta^*) = \epsilon(\theta)+ 1 \   \ \ (mod\ 2)\,, \\
&&gh(\theta^*) = -gh(\theta) - 1\ .
\eearst
Thus in particular we have that $gh(\eta) = - gh(\bar{\cP}) = 1;\ gh(q^*) = gh(p^*) = -1;\ gh(\eta^*) = -2;\ gh(\bar{\cP}^*) = 0$.

Viewed as a function of the above variables, the unitary BFV-action (\ref{13.BV-action}) takes the form (we denote it now by $S_H$)
\bear
S_{H}[q,p,\eta,\bar{\cal P};q^*,p^*,\eta^*,{\bar{\cP}}^*]&=&\int dt \left(\dot q^ip_i + \dot\eta^A{\bar{\cP}}_A 
- {H}_{B}(q,p,\eta,\bar{\cal P})\right)\nonumber\\
&+& \int dt \left(q^*_i \delta_B q^i + p^*_i\delta_B p_i+\eta^*_A \delta_B\eta^A
+{\bar{\cP}}^*_A \delta_B{\bar{\cP}}_A\right)\nonumber
\eear
or
\be\label{13.S_Hstar(1)}
S_H = \int dt \left({\dot q}^i p_i + {\dot\eta}^A{\bar{\cal P}}_A - {H}_{B}(q,p,\eta,{\bar{\cal P}})\right)
+\int dt\ \theta^*_\ell\delta_{B}\theta^\ell\ .\nonumber
\ee
In a fixed gauge, i.e. for a given $\Psi_H$ in (\ref{13.def.antifield(2)}), the variables $\{\theta^*_{\ell}\}$ are functions of $\{\theta^\ell\}$. On the other hand, if the action is regarded as a function of the independent variables 
$q^i$, $p_i$, $\eta^A$ and 
${\bar{\cal P}}_A$ and their ``starred" counterparts $q^*_i$, $p^{*i}$, $\eta^*_A$ and ${\bar{\cal P}}^{*A}$,
then - as we now show - (\ref{13.S_Hstar(1)}) is found to satisfy the (phase-space) master equation,
\be\label{13.Hamiltonian-master-equation}
(S_H,S_H) = 0\,,
\ee
with the ``antibracket" (f,g) defined as follows,
\be\label{13.antibracket}
(f,g) \equiv \sum_\ell\int dt\ \left(\frac{\delta^{(r)} f}{\delta\theta^\ell(t)}\frac{\delta^{(l)} g}{\delta\theta^*_\ell(t)}
-\frac{\delta^{(r)} f}{\delta\theta^*_\ell(t)}\frac{\delta^{(l)} g}{\delta\theta^\ell(t)}\right)
\ .
\ee
Here $\ell$ labels the phase space degrees of freedom (which in the case of fields will also include the spacial coordinates). In the case of a continuous index, the sum is understood to be an integral. 

We now verify that (\ref{13.S_Hstar(1)}) is a solution to the master equation (\ref{13.Hamiltonian-master-equation}).
\footnote{See [20] for an alternative proof using functional methods}
The proof is based on the BRST invariance of the first integral in (\ref{13.S_Hstar(1)}) and the nilpotency of $Q_{\cB}$ which, together with the (graded) Jacobi identity implies that
$\{Q_{\cB}\{Q_{\cB},\Psi\}\} = 0$.

Consider the antibracket
\be\label{13.(S_H,S_H)(1)} 
(S_{H},S_{H}) = \sum_k\int dt\ \left(\frac{\delta^{(r)}S_H}{\delta\theta^\ell(t)}\frac{\delta^{(l)} 
S_H}{\delta\theta^*_\ell(t)}
-\frac{\delta^{(r)}S_H}{\delta\theta^*_\ell(t)}\frac{\delta^{(l)}S_H}
{\delta\theta^\ell(t)}\right)\ .
\ee 
Since $S_H$ is a Grassmann even operator, and either $\theta^\ell$ or $\theta^*_\ell$ has non-vanishing Grassmann signature, we have that
\footnote{We suppress the time argument of $\theta_\ell$ from now on.}
\[ 
\frac{\delta^{(r)}{S_H}}{\delta\theta^*_\ell}\frac{\delta^{(l)}{S_H}}{\delta\theta^\ell} 
= -\frac{\delta^{(r)}{S_H}}{\delta\theta^\ell}\frac{\delta^{(l)}{S_H}}
{\partial\theta^*_\ell}\ .
\]
Hence (\ref{13.(S_H,S_H)(1)}) takes the simpler form
\be\label{13.(S_H,S_H)(2)}
({S_H},{S_H}) = 2\sum_k\int  dt\ \frac{\delta^{(r)}{S_H}}{\delta\theta^k}\frac{\delta^{(l)} 
{S_H}}{\delta\theta^*_k} \ .
\ee
Let us now decompose ${S_H}$ in (\ref{13.S_Hstar(1)}) as follows:
\[
S_H = S_B[\theta] + \Delta[\theta,\theta^*] \ ,
\]
where
\[ 
S_B = \int dt\left({\dot q^i}p_i
+ {\dot\eta}^A{\bar{\cal P}}_A - {H}_{B}(q,p,\eta,{\bar{\cal P}})\right) 
\]
is a $BRST$ invariant action and
\be\label{13.Delta} 
\Delta = \sum_\ell\int dt\ \theta^*_\ell\{Q_{\cB},\theta_\ell\}\ .
\ee
Since
\be\label{13.delta_Btheta} 
\frac{\delta^{(l)}S_H}{\delta\theta^*_\ell} = \delta_B\theta^\ell\,,
\ee
with $\delta_{B}\theta$ the BRST variation (\ref{13.delta_B}),
we have that (\ref{13.(S_H,S_H)(2)}) is given by
\bearst 
(S_H,S_H) &=& 2\sum_\ell \int dt \frac{\delta^{(r)} S_H}{\delta\theta^\ell}\delta_{B}\theta^\ell\\
&=& 2\sum_\ell\int dt\ \frac{\delta^{(r)} S_B}{\delta\theta^\ell}\delta_B\theta^\ell + 
2\sum_\ell\int dt\ \frac{\delta^{(r)}\Delta}{\delta\theta^\ell}\delta_B\theta^\ell\ .
\eearst
The first sum vanishes since $S_{B}$ is $BRST$ invariant. The second integral, is just twice the $BRST$ variation of (\ref{13.Delta}) with respect to $\theta^\ell$, with
\[
\delta_B\Delta = \int dt\ \theta^*_\ell\Bigl(\delta_B\{Q_{\cB},\theta_\ell\}\Bigr) = 
\int dt\ \theta^*_\ell\{Q_{\cB},\{Q_{\cB},\theta_\ell\}\}\ .
\]
Making use of the Jacobi identity
\[
(-1)^{\epsilon_h \epsilon_g }\{h,\{f,g\}\} + cyclic\ perm. = 0\,,
\]
this term is seen to vanish because of the nilpotency of $Q_{\cal B}$, i.e.,
$\{Q_{\cB},Q_{\cB}\} = 0$. 

We have thus shown that the $BRST$ invariance of $S_B$ and the nilpotency of $Q_{\cB}$ imply that the action $S_{H}$, considered as a function of the fields and antifields, satisfies the 
hamiltonian master equation (\ref{13.Hamiltonian-master-equation}). Note that for this to be the case, 
the linear dependence of $S_H$ on the antifields, which is inherent to the hamiltonian formulation, was important.

From (\ref{13.delta_Btheta}) and the definition of the antibracket (\ref{13.antibracket})
we see that we also have
\be\label{13.delta_Htheta}
\delta_B\theta^\ell = (\theta^\ell,S_H) \equiv \delta_H\theta^\ell\,,
\ee
so that $S_H$ can be regarded as the generator of $BRST$ transformations in the antibracket sense.
It follows in particular that the $BRST$ invariant (gauge-fixed) unitarized BFV action
\[
S_U[\theta] = S_H[\theta,\frac{\delta\Psi_H}{\delta\theta}]
\]
satisfies
\[
(S_U,S_H) = 0\,.
\]
Extending the variation (\ref{13.delta_Htheta}) to include the antifields,
\be\label{13.delta_Hthetastar}
\delta_H\theta^*_{\ell} = (\theta^*_{\ell},S_H)\,,
\ee
it is clear from (\ref{13.Hamiltonian-master-equation}) that the complete set of transformations is also a symmetry of $S_H[\theta,\theta^*]$.

\section{Transition to configuration space.}

Consider the BFV gauge fixed phase space partition function $Z_\Psi$, which we write in the form,
\bear
Z_{\Psi} &=& \int {D}q {D}p{D}\eta{D}{\bar{\cal P}}
\int {D}q^* {D}p^*{D}\eta^*{D}{\bar{\cal P}}^*
\prod_{i,t}\delta\left(q^*_i + \frac{\delta \Psi_H}{\delta  q^i}\right)
 \delta\left(p^{*i} + \frac{\delta \Psi_H}{\delta p_i}\right)\nonumber\\
&\times&\prod_{A,t}\delta\left(\eta^*_A - \frac{\delta^{(r)} \Psi_H}{\delta\eta^A}\right)\delta\left({\bar{\cal P}}^{*A} - \frac{\delta^{(r)} \Psi_H}{\delta{\bar{\cal P}}_A}\right)e^{iS_H[q,\eta,{\bar{\cal P}};q^*,\eta^*,{\bar{\cal P}}^*]} \label{13.Z_Psi(1)}\,, 
\eear
where $S_H$ is given by (\ref{13.S_Hstar(1)}).
The transition to the lagrangian formulation cannot be effected generically. Assumptions must be made, which however include many cases of physical interest. In particular we will consider  gauge theories with only first class constraints, and of unit rank. For such systems the algebra of gauge transformations close off shell. Furthermore we consider systems 
where each primary constraint gives rise to just one secondary constraint. Hence the number of gauge identities, and therefore also gauge parameters, equals the number of primaries (or secondaries).
\footnote{This seems to have been also implicitely assumed in the work of BFV [6-10].}
For systems of rank one the BRST Hamiltonian and BRST charge take a particular simple form. To keep the discussion as simple and transparent as possible, we will, for the moment, consider systems with a finite number of degrees of freedom. The extension to systems with an infinite number of degrees of freedom is then obvious.

Suppose we are given the classical canonical Hamiltonian $H_c(q,p)$ and the first class constraints $G_A$. The primary constraints we denote by $\phi_\alpha$ ($\alpha = 1,\cdots,r$). In the following it is convenient to define the secondary constraints $T_\alpha$ by the strong equality
\be\label{13.def.secondary-constraints}
\{\phi_\alpha,H_c\} = T_\alpha \  \ \ (secondary\ constraints)\ .
\ee
We denote the set of constraints collectively by $\{G_A\} := (\{\phi_\alpha\},\{T_\alpha\})$. 
For a gauge theory of rank {\it one} the $BRST$ charge and Hamiltonian are then given by [8]
\be\label{13.rank1-BRST-charge}
Q_{\cB} = G_A\eta^A + \half{\bar{\cP}}_A U^{A}_{BC}\eta^B\eta^C \,,
\ee
and
\be\label{13.rank1-BRST-Hamiltonian}
{H}_{B}(q,p,\eta,{\bar{\cP}}) = {H}_c(q,p) + {\bar{\cP}}_AV^A_{\ B}{\,\,}\eta^B \,,
\ee
where $U^A_{BC}$ and $V^A_{\ B}$ are the structure functions defined in (\ref{13.GG-PB}) and (\ref{13.HG-PB}).
In terms of the ``Faddeev-Popov ghosts" $c^\alpha$ and antighosts ${\bar c}_\alpha$, and their conjugate momenta ${\bar P}_\alpha$ and ${P}^\alpha$ ($\alpha = 1,\cdot\cdot\cdot,r$), $\eta^A$ and ${\bar P}_A$ are 
given in vector form by [6]
\be
\vec\eta= 
\left(\begin{array}{c} \ \ 
\vec P\\
\vec c
\end{array}\right)\label{13.eta}
\ee
and
\be
\vec{\bar{\cal P}}_A := 
\left(\begin{array}{c}
\vec{\bar c}\\
\vec{\bar P}
\end{array}\right)\label{13.cP}\ .
\ee
A generic transition to configuration space requires one to perform all momentum integrations. We will thus make a number of simplifying assumptions, which can however be relaxed in particular cases:

\bigskip
i) The constraints $G_A(q,p)$ are at most linear in the momenta $p_i$.

\bigskip
ii) The structure functions $V^A_{\ B}$ and $U^C_{AB}$ are of the form 
\be\label{13.YM-VU-matrix}
V^B_{\ A} =\left(\begin{array}{cc}
\undertilde 0&\undertilde 0\\
\delta^\beta_{\ \alpha} &h^\beta_{\ \alpha}
\end{array}\right) \ .\quad
U^C_{AB}: =\left(\begin{array}{cc}
\undertilde 0&\undertilde 0\\
\underline 0&\kappa^\gamma_{\alpha\beta}
\end{array}\right) \,,
\ee 

and do not depend on the momenta $p_i$. 

\bigskip
iii) The ``fermion gauge fixing functional" $\Psi_H$ is independent of the momenta 
 $p_i$, and the ghost momenta $P^A$, ${\bar P}_A$ in (\ref{13.eta}) and (\ref{13.cP}).

\bigskip\noindent
Assumption ii) implies a) that the primary constraints are in strong involution with all the constraints; b) the algebra of the secondaries closes on itself, and c) that the
algebra of the secondary constraints with the Hamiltonian is a linear combination of the secondaries. The entry $+\delta^\alpha_{\ \beta}$ in $V^A_{\ B}$ follows directly from the definition of the secondary constraint (\ref{13.def.secondary-constraints}). These properties are realized e.g. 
in the case of $SU(N)$ gauge theories and the bosonic string.

Note that iii) is not really a restriction. In fact, since
by the Fradkin-Vilkovisky theorem [26] the partition function does not depend on the choice of gauge (i.e., it is independent of $\Psi_H$), we are completely free to chose 
$\Psi_H$ in a convenient way. This will of course also correspond to a particular choice of the "lagrangian gauge fixing function" $\Psi_L$ in the field-antifield action, where the partition function does also not depend on $\Psi_L$ [14,22].

\bigskip\noindent
An immediate consequence of iii) is that,
\[ 
p^*_i = P^*_\alpha = {\bar P}^{*\alpha} = 0 \ .
\]
Because of ii) the $BRST$ Hamiltonian (\ref{13.rank1-BRST-Hamiltonian}) and $BRST$ charge
(\ref{13.rank1-BRST-charge}) take the form
\be\label{13.rank1H_B}
H_{B} = H_c + {\bar P}_\alpha P^\alpha + {\bar P}_\alpha h^\alpha_{\ \beta}(q) c^\beta \ ,
\ee
\be\label{13.rank1Q_B}
Q_{\cB} = P^\alpha\phi_\alpha + c^\alpha T_\alpha + \frac{1}{2}\kappa^\gamma_{\alpha\beta}
{\bar P}_\gamma c^\alpha c^\beta \ .
\ee
From (\ref{13.rank1Q_B}) and the definition of the graded Poisson brackets (\ref{13.PB1}), it then follows that 
\be\label{13.delta-cbar} 
\{Q_{\cB},\bar c_\alpha\} = -\phi_\alpha \ ,
\ee
where $\phi_\alpha = 0$ are the primary constraints. In the following we assume that the $\phi_\alpha$'s are given by
\footnote{We thereby assume that for a suitable choice of coordinates the lagrangian is independent of the velocities $\dot q^\alpha$.}
\[ 
\phi_\alpha = p_\alpha \ \ ; \ \ (\alpha = 1,\dots,r)\ .
\]
The $T_\alpha$'s, defined by (\ref{13.def.secondary-constraints}), then do not depend on the $p_\alpha$'s, since the canonical Hamiltonian $H_c$ is defined on the primary surface.

Let us separate the phase space variables $(q,p)$ into the dynamical ones, which we label as $(q^a,p_a), a = 1,\cdots ,n-r$, 
and the remaining non-dynamical variables, $(q^\alpha,p_\alpha), \alpha = 1,\cdots ,r$.  
According to (\ref{13.delta_B}) the BRST-charge (\ref{13.rank1Q_B}) induces the following  transformations:
\be\label{13.delta_B-transf}
\delta_B q^a = -c^\alpha\frac{\partial T_\alpha}{\partial p_a}\ , \quad
\delta_B c^\alpha = -\frac{1}{2}\kappa^\alpha_{\,\beta\gamma}c^\beta c^\gamma\ 
\ee
\be\label{13.delta-bar-c} 
\delta_B{\bar c}_\alpha = -\phi_\alpha = -p_\alpha\ , 
\ee
and
\be\label{13.deltaq^alpha}
\delta_Bq^\alpha = -P^\alpha \,.
\ee 
Performing the integration over  ${\bar P}_\alpha$ in (\ref{13.Z_Psi(1)})  yields a $\delta$-function which fixes $P^\alpha$ as a function of the coordinates and their time derivatives (pull-back of the Legendre transformation) 
\[
P^\alpha \to -D^\alpha_{0\beta}(q)c^\beta \ ,
\]
where
\be\label{13.cov-derivative}
D^\alpha_{0\beta}(q) = \delta^\alpha_\beta\partial_0 + h^\alpha_{\ \beta}(q) \ .
\ee
The BRST transformations (\ref{13.delta_B-transf}) and (\ref{13.deltaq^alpha}) are then replaced by
\bear\label{13.pullback(1)}
\hat\delta q^a &=& -c^\alpha\frac{\partial T_\alpha}{\partial p_a} ,\\
\hat\delta c^\alpha &=& -\frac{1}{2}\kappa^\alpha_{\beta\gamma}c^\beta c^\gamma\\
\hat\delta q^\alpha &=& D^\alpha_{0\beta}(q)c^\beta \,,
\eear
Carrying out the integration over $\{p_a\}$,
\footnote{Recall that $\Psi$ is assumed to be at most a function of $c$, $\bar c$ and $q$.}
we are left with the following expression for the partition function in configuration space,
\bear
Z^L_\Psi &=& \int \prod_\alpha Dp_\alpha\int Dq Dc D\bar c \int Dq^* Dc^* D{\bar c}^*\prod_{i,t}\delta\left(q^*_i + \frac{\delta \Psi_H}{\delta q_i}\right)\nonumber\\
&&\times\prod_{\alpha,t}\delta\left(c^*_\alpha - \frac{\delta^{(r)} \Psi_H}{\delta c^\alpha}\right)
\times\prod_{\alpha,t}\delta\left({\bar c}^{*\alpha} - \frac{\delta^{(r)} \Psi_H}{\delta {\bar c}_\alpha}\right)e^{i\int dt\ L(q,\dot q)}\label{13.Z^L_Psi(2)}\\ 
&&\times e^{i\int dt\,\{q^*_\alpha - {\dot{\bar c}}_\alpha)\hat\delta q^\alpha + 
q^*_a{\hat\delta} q^a + c^*_\alpha\hat\delta c^\alpha - (\bar{c}^{\alpha*} -\dot q^\alpha)p_\alpha\}}\,,\nonumber 
\eear
where we have made use of (\ref{13.delta-bar-c}). 
Note that we have now written $\hat\delta$ instead of $\delta_B$ in (\ref{13.Z^L_Psi(2)}), since after carrying out the momentum integrations, $\hat\delta\theta^\ell$ 
can in general not be identified with $\delta_B\theta^\ell$, but is a function of the coordinates $\theta^\ell$ and their time derivatives, obtained after performing the ``pullback" to 
configuration space.

The ``star" variables in (\ref{13.Z^L_Psi(2)}) are so far defined in terms of Hamiltonian functional $\Psi_H$, which was conveniently chosen not to depend on the momenta. 
Let us now introduce the following ``{\it lagrangian} fermion gauge fixing functional"
\be\label{13.Psi_L} 
\Psi_L[q,c,\bar c] = \Psi_H[q,c,\bar c]-\int dt\ {\bar c}_\alpha\dot q^\alpha \ .
\ee
Then (\ref{13.Z^L_Psi(2)}) can be written in the form
\bear
Z^L_\Psi &=&\int DB\int Dq Dc D\bar c \int Dq^*Dc^*D{\bar c}^*
\prod_{i,t}\delta\left(q^*_i + \frac{\delta \Psi_L}{\delta q^i}\right)
\prod_{\alpha,t}\delta\left({c}^*_\alpha - \frac{\delta^{(r)} \Psi_L}
{\delta c^\alpha} \right)\nonumber\\
&\times&\prod_{\alpha,t}\delta\left({\bar c}^{*\alpha} 
- \frac{\delta^{(r)} \Psi_L}{\delta {\bar c}_\alpha}\right)
e^{i\int dt\,S[q,c,\bar c;\,q^*,c^*,{\bar c}^*,B]}\label{13.Z^L_Psi(3)}
\eear
with
\be\label{13.Sstar}
S[q,c,\bar c;q^*,c^*,{\bar c}^*,B] =S_{cl}[q] +\int dt\, \left( q^*_i\hat{\delta} q_i + c^*_\alpha{\hat\delta}c^\alpha+{\bar{c}}^{*\alpha} B_\alpha\right) \ ,
\ee
where the ``star" variables (antifields) are now fixed in terms of the functional derivatives of (\ref{13.Psi_L}) in a completely analogous way as in the hamiltonian BFV formulation 
(\ref{13.def.antifield(2)}),
and where we have set $B_\alpha = -p_\alpha$. 
The last contribution in (\ref{13.Sstar}) is a term which in the axiomatic approach is usually introduced as a non-minimal trivial extension to the field-antifield action [24],
which does not manifest itself in the master equation. Here it is seen to be generated from the BRST exact term in the BFV phase-space formulation.
Note that the dependence on $\bar c$ (Faddeev-Popov antighosts) comes in implicitely via the shift (\ref{13.Psi_L}). 
 
Upon carrying out the integrations over the antifields, which now become fixed functions of $q_i, c^\alpha$, and $\bar c_\alpha$, and choosing $\Psi_H$ in (\ref{13.Psi_L})
to be
\footnote{Note that if we assume $\Psi_H$ to be linear in the antighosts ${\bar c}_\alpha$, then this is the most general choice, since $\Psi_H$ must have ghost number $gh(\Psi_H) = -1$, and was assumed not to depend on the momenta.}
\be\label{13.chi}
\Psi_H = \int dt\, {\bar c}_\alpha \chi^\alpha[q(t)]\,,
\ee
we have that $c^*_\alpha = 0$, ${\bar c}^{*\alpha} = -{\dot q}^\alpha + \chi^\alpha$ in 
(\ref{13.Sstar}), so that $B_\alpha$ 
in (\ref{13.Sstar}) is seen to play the role of a Lagrange multiplier implementing the gauge ${\dot q}^\alpha - \chi^\alpha=0$. Note that for the gauge considered here, where the fermion gauge fixing function
does not depend on the ghosts $c^\alpha$, we have $c^*_\alpha = 0$, so that only the knowledge of
the symmetry of the {\it classical} action is required for constructing $Z^L_\Psi$.

Under the assumptions stated above, we see that, analogous to the hamiltonian BFV action 
(\ref{13.S_Hstar(1)}), the lagrangian action obtained above
is again a {\it linear} function of the antifields. We want to emphasize once more that, although we have made a particular convenient choice of gauge, we have actually proven the equivalence 
of the hamiltonian BFV formulation and lagrangian BV-formulation, since both partition functions do not depend on the respective "fermion gauge fixing function". 

\bigskip\noindent
We now consider an example.

\bigskip\noindent
{\it The Yang-Mills theory}

\bigskip\noindent
The $SU(3)$ Yang-Mills theory is an example satisfying all the assumptions made above.  
Its unitarized BFV phase-space action is given by 
\[ 
S_U =  \int d^4x \Bigl(\dot A^\mu_a\pi^a_\mu + \dot\eta^A{\bar{\cP}}_A - {\cal H}_c - {\bar{\cal P}}_AV^A_{\ B}\eta^B + \{\Psi,Q_{\cB}\}\Bigr)
\]
where  
\[
{\cal H}_c = \frac{1}{2}\pi^a_i\pi^a_i + \frac{1}{4}F^a_{ij}F^a_{ij}- A^a_0D^i_{ab}\pi^b_i
\]
is the canonical hamiltonian density evaluated on the primary constraint surface, $D^i_{ab}$ is the covariant derivative, and 
\[
Q_{\cB} = \int d^4x\Bigl(\eta^AG_A + \frac{1}{2}f_{abc}{\bar P}_ac^bc^c\Bigr) 
= \int d^4x\left(P^a\pi^a_0+c^aT_a + \frac{1}{2}f_{abc}{\bar P}_ac^bc^c\right)\ .
\]
is the BRST charge which is of the form (\ref{13.rank1Q_B}); $a,b,\cdots$ are the $SU(3)$ color indices, $f_{abc}$ are the structure constants of $SU(3)$, and $G_A=0$, with $G_A :\equiv (\{\phi_a\},\{T_a\})$, is the set of primary and secondary constraints. Here $\phi_a = \pi^a_0$ and $T_a(x) = D^i_{ab}\pi^b_i$. The matrices $V^A_{\ B}$ and $U^C_{AB}$ have the form (\ref{13.YM-VU-matrix}) with
$h^\alpha_{\ \beta} \to -gf_{abc}A^c_0(x)$ and $\kappa^\gamma_{\alpha\beta} \to 
f_{abc}$ .
We can now immediately translate the partition function (\ref{13.Z^L_Psi(3)}) to the case of the Yang-Mills theory:
\bearst 
Z^L_{YM} &=& \int {\cD}B\int {\cD}A{\cD}c{\cD}{\bar c}\prod_{i,a}\left(A^{i*}_a(x)+\frac{\partial\Psi_L(x)}
{\partial A^a_i(x)}\right) 
\prod_a\left(c^*_a(x)-\frac{\partial\Psi_L(x)}{\partial c^a(x)}\right)\nonumber \\
&&\prod_a\left({\bar c}^{*a}(x)-\frac{\partial\Psi_L(x)}{\partial{\bar c}_a(x)}\right)
e^{iS_L[A,c,\bar c;A^*,c^*,{\bar c}^*;B]}\label{11.Z^L_{YM}(1)} \ .
\eearst
where
\be\label{13.YM:Sstar}
S_L = \int d^4x\ \left(-\frac{1}{4}F^{\mu\nu}_aF^a_{\mu\nu} + A^{*i}_a\hat\delta A_i^a + A^{*0}_a\hat\delta A^a_0 + c^*_a{\hat\delta}c^a + {\bar c}^{*a}B_a\right)
\ee
is the field-antifield action. Although $\Psi_L$ in (\ref{13.Z^L_Psi(3)}) had actually been chosen to be of the form (\ref{13.Psi_L}), the partition function does not depend on $\Psi_L$. Hence we do not need to specify it at this point.
Translating (\ref{13.pullback(1)}) and (\ref{13.cov-derivative}) to the case of the Yang-Mills theory yields
\[ 
\hat\delta A^a_\mu = (D_\mu)_{ab}c^b\ ,
\]
where $(D_\mu)_{ab}$ is the the covariant derivative. 
The expressions for the remaining variations can be read off directly from the corresponding Poisson brackets with the $BRST$ charge, analogous to (\ref{13.delta_B}), since they do not involve the momenta which have been integrated out:
\bearst 
&&\hat\delta c^a(x) = -\frac{g}{2}f_{abc}c^b(x)c^c(x)\ ,\\
&&\hat\delta{\bar c}^a = B^a\ ,\\
&&\hat\delta B^a = 0\ ,
\eearst
where we have set $B^a = -\pi^a_0$. Notice that these are the well known 
expressions implementing the $BRST$ symmetry of the gauge fixed action on the lagrangian level. 
Inserting these expressions in (\ref{13.YM:Sstar}) one is led to
\[ 
S = \int d^4x\ \Bigl(-\frac{1}{2}F^a_{\mu\nu}F^{\mu\nu}_a + {A^a_\mu}^*{D}^\mu_{ab}c^b 
-\frac{g}{2}f_{abc}c^*_ac^bc^c + {\bar c}^{*a}B_a\Bigr)\,.
\]
Note that the non-abelian structure of the YM-theory has induced a term binlinear in the 
ghost fields, carrying vanishing ghost number and Grassmann parity. We now choose 
\[ 
\Psi_L = \int d^4x \ {\bar c}_a \partial^\mu A^a_\mu(x)\,.
\]
This $\Psi_L$ will implement the Lorenz gauge, since ${\bar c}^*_\alpha$ in (\ref{13.YM:Sstar}) 
becomes
\[
{\bar c}^{*a}(x) = \partial^\mu A^a_\mu(x)\ .
\]
The remaining antifields are fixed as follows:
\bearst
&&{A^a_\mu}^*(x) = -\partial^\mu{\bar c}_a(x)\ ,\\
&&c^*_a(x) = 0\ .
\eearst
At this stage the antighosts have made their 
appearance, while the ghosts $c^a$ were present already before fixing the gauge. With the above choice of $\Psi_L$ we have in particular $c^*_a=0$. One is then led to the familiar Faddeev-Popov result for the partition function in the Lorentz gauge.

\section{ The lagrangian master equation}

We have seen that in a semi-classical framework, the hamiltonian master equation (\ref{13.Hamiltonian-master-equation}) is satisfied
in general, without any restriction. This was a consequence of the linear dependence of
the hamiltonian action on the ``star" variables. This is no longer necessarily true on the lagrangian level,
where $S$ can in general depend on higher powers of the antifields. However, as we now show, the field-antifield action (\ref{13.Sstar}) satisfies the master equation.

Motivated by the hamiltonian derivation, one is naturally led to introduce
the lagrangian antibracket \cite{BV1},
\[
(f,g) \equiv \sum_\ell \int dt\,\left(\frac{\delta^{(r)} f}{\delta\vartheta^\ell(t)}\frac{\delta^{(l)} g}{\delta\vartheta^*_\ell(t)}
-\frac{\delta^{(r)} f}{\delta\vartheta^*_\ell(t)}\frac{\delta^{(l)} g}{\delta\vartheta^\ell(t)}\right)\,,
\]
where now $\{\vartheta_\ell\} = \{q_i,c^\alpha,\bar c_\alpha\}$. For the systems of {\it rank one} in question, where $S$ is a linear function of the star-variables, we have that $\hat\delta\theta^\ell = (\theta^\ell,S)$, where $\hat\delta\theta^\ell$ is the pullback of the BRST variation $\delta_B\theta^\ell$. We therefore have the correspondence
\[
\hat{\delta}\vartheta^\ell = (\vartheta^\ell,S) \leftrightarrow \delta_{B}\theta^\ell = 
\{Q_{\cB},\vartheta^\ell\}\ .
\]
More general we have for a function $f$ of the $\vartheta$-variables,
\[
\hat{\delta}f(\vartheta) = (f(\vartheta),S) \leftrightarrow \delta_{B}f(\vartheta) = 
\{Q_{\cB},f(\vartheta)\}\ .
\]
In particular, the nilpotency of $\delta_B$ implies the nilpotency of $\hat\delta$,
since the latter is the pullback of the former. Hence
\[
{\hat{\delta}}^2f(\vartheta) = ((f(\vartheta),S),S)= \half ((S,S),f) = 0\,,
\]
where we made use of the Jacobi identity
\[
(-1)^{(\epsilon_f + 1)(\epsilon_h + 1)}((f,g),h) + cyclic\,\,perm. = 0.
\]
Since $(S,S)$ is a linear function of the antifields, and $f$ is arbitrary, we conclude from here that
\be\label{13.Lagrangean-master-equation} 
(S,S) = 0\,.
\ee
We have thus arrived at the so called ``classical" master equation. 

Finally, as in the case of (\ref{13.delta_Hthetastar}), we extend these transformation laws to include the antifields:
\be\label{13.deltaThetaStar}
\hat\delta\vartheta^*_\ell = (\vartheta^*_\ell,S)\,.
\ee
Regarding $S$ as the generator of generalized $BRST$ transformations, does not only imply the invariance of
$S$ under the transformations generated by $S$, but also the invariance of the gauge fixed action 
\[
S_{eff}=S[\vartheta,\vartheta^*]_{\vartheta^*=-\epsilon(\vartheta)\frac{\delta\Psi_L}
{\delta\vartheta}}. 
\]
Indeed, consider (the summation also includes an integral over time)
\[
(S_{eff},S) = \sum_\ell\frac{\delta^{(r)}S_{eff}}{\delta\vartheta^\ell}\frac{\delta^{(\ell)}S}{\delta\vartheta^*_\ell}\ .
\]
Now
\[ 
\frac{\delta^{(r)}S_{eff}}{\delta\vartheta^\ell} = \left(\frac{\delta^{(r)}S}{\delta\vartheta^\ell}\right)_\Sigma + 
\sum_k\left(\frac{\delta^{(r)}S}{\delta\vartheta^*_k}\right)
\frac{\delta^{(r)}}{\delta\vartheta^\ell}\left(\frac{\delta^{(r)}\Psi_L}{\delta\vartheta^k}\right)\ ,
\]
where $\Sigma$ is the surface 
\[
\Sigma : \ \ \vartheta^*_\ell -\frac{\delta^{(r)}\Psi_L}{\delta\vartheta^\ell} = 0\ .
\]
Hence
\be\label{13.(S_eff,S)} 
(S_{eff},S) = \half(S,S)_\Sigma + \sum_{\ell,k}\left(\frac{\delta^{(r)}S}{\delta\vartheta^*_\ell}\right) K_{\ell k} 
\left(\frac{\delta^{(\ell)}S}{\delta\vartheta^*_k}\right)\ ,
\ee
where
\[ 
K_{\ell k} = \frac{\delta^{(r)}}{\delta\theta^\ell}\left(\frac{\delta^{(r)}\Psi_L}
{\delta\vartheta^k}\right)\ .
\]
The second term on the RHS of (\ref{13.(S_eff,S)}) can be shown to vanish by making 
use of the Grassmann properties of the products and the relation between right-and-left derivatives. We thus finally arrive at the statement that
\[ 
(S_{eff},S) = \half(S,S)_\Sigma\,,
\]
Hence if 
(\ref{13.Lagrangean-master-equation}) holds, then $S_{eff}$ is invariant under transformations generated by $S$. .

\section{Conclusion}

Starting from the Batalin-Fradkin-Vilkovisky (BFV) BRST invariant partition function, we have shown -for the case of first rank gauge theories with momentum independent structure functions- that the ``antifields" and gauge-fixing conditions of the Batalin-Vilkovisky lagrangian formulation emerge in a very natural way.
In particular our phase space approach leads automatically to the non-minimally extended
BV lagrangian action, with the Lagrange multipliers usually introduced to implement a particular gauge identified with the primaries of the {\it classical} lagrangian. We have also have given a short proof that the field-antifield action thus obtained solves the master equation.

\end{document}